\documentclass[sigconf]{acmart}
\PassOptionsToPackage{table,xcdraw}{xcolor} 
\usepackage{colortbl}
\usepackage{tcolorbox} 

\usepackage{tikz}
\usepackage{mathdots}
\usepackage{cancel}
\usepackage{siunitx}
\usepackage{array}
\usepackage{multirow}
\usepackage{gensymb}
\usepackage{tabularx}
\usepackage{extarrows}
\usepackage{booktabs}
\usetikzlibrary{fadings}
\usetikzlibrary{patterns}
\usetikzlibrary{shadows.blur}
\usetikzlibrary{shapes}
\usepackage{listings}
\usepackage{array}
\usepackage{geometry}
\usepackage{caption}
\usepackage{textcomp}

\newcommand{\lstbg}[3][0pt]{{\fboxsep#1\colorbox{#2}{\strut #3}}}
\lstdefinelanguage{diff}{
  basicstyle=\footnotesize \ttfamily \color{black},
  columns=fullflexible,
  breaklines=true,
  breakatwhitespace=false,
  showspaces=false,               
  showstringspaces=false,  
  frame=single, 
  showtabs=false,
  numbersep=5pt,
  showstringspaces=false,        
  stepnumber=1,                   
  tabsize=5,                     
  title=\lstname,  
  numbers=left,                 
  numbersep=5pt,   
  backgroundcolor=\color{white},
  morecomment=[f][\lstbg{red!30}]-,
  morecomment=[f][\lstbg{green!30}]+,
  morecomment=[f][\textit]{@@},
  moredelim=**[is][\color{red}]{@}{@},
  moredelim=**[is][\color{green}]{##}{##},
}
\definecolor{white}{rgb}{0.98,0.98,0.98}
\definecolor{dkgreen}{rgb}{0,0.6,0}
\definecolor{dred}{rgb}{0.545,0,0}
\definecolor{dblue}{rgb}{0,0,0.545}
\definecolor{lgrey}{rgb}{255,0.9,0.9}
\definecolor{gray}{rgb}{0.4,0.4,0.4}
\definecolor{darkblue}{rgb}{0.0,0.0,0.6}

\definecolor{pblue}{rgb}{0.13,0.13,1}
\definecolor{pgreen}{rgb}{0,0.5,0}
\definecolor{pred}{rgb}{0.9,0,0}
\definecolor{pgrey}{rgb}{0.46,0.45,0.48}
\lstdefinestyle{prStyle}{
  showspaces=false,
  showtabs=false,
  breaklines=true,
  showstringspaces=false,
  breakatwhitespace=true,
  postbreak=\mbox{\textcolor{black}{$\hookrightarrow$}\space},
  captionpos=b,    
  commentstyle=\color{pgreen},
  keywordstyle=\color{pblue},
  stringstyle=\color{pred},
  basicstyle=\ttfamily,
  moredelim=[il][\textcolor{pgrey}]{},
  frame=tb,
  moredelim=[is][\textcolor{pgrey}]{\%\%}{\%\%}
  moredelim=**[is][\color{red}]{@}{@},
  moredelim=**[is][\color{dkgreen}]{##}{##},
}

\lstdefinelanguage{cpp}{
      backgroundcolor=\color{white},  
      basicstyle=\scriptsize \ttfamily \color{black} ,   
      breakatwhitespace=false,       
      breaklines=true,
      postbreak=\mbox{\textcolor{black}{$\hookrightarrow$}\space},
      captionpos=b,                   
      commentstyle=\color{dkgreen},   
      deletekeywords={...},          
      escapeinside={\%*}{*)},                  
      frame=single,                  
      language=C++,                
      keywordstyle=\color{dblue},  
      morekeywords={BRIEFDescriptorConfig,string,TiXmlNode,DetectorDescriptorConfigContainer,istringstream,cerr,exit}, 
      identifierstyle=\color{black},
      stringstyle=\color{blue},      
      numbers=right,                 
      numbersep=5pt,                  
      numberstyle=\tiny\color{black}, 
      rulecolor=\color{black},        
      showspaces=false,               
      showstringspaces=false,        
      showtabs=false,                
      stepnumber=1,                   
      tabsize=5,                     
      title=\lstname,
      moredelim=**[is][\color{red}]{@}{@},
      moredelim=**[is][\color{dkgreen}]{##}{##},
    }

\AtBeginDocument{%
  }

\setcopyright{acmlicensed}
\copyrightyear{2018}
\acmYear{2018}
\acmDOI{XXXXXXX.XXXXXXX}

\usetikzlibrary{shapes.geometric, arrows.meta, positioning}

\acmConference[Conference acronym 'XX]{Make sure to enter the correct
  conference title from your rights confirmation emai}{June 03--05,
  2018}{Woodstock, NY}

\acmISBN{978-1-4503-XXXX-X/18/06}

\newcommand{\tool}[0]{\mbox{\textsc{LibProf}}}

\renewcommand\footnotetextcopyrightpermission[1]{} 

\begin{document}

\title{\tool{}: A Python Profiler for Improving Cold Start Performance in Serverless Applications}

\author{Syed Salauddin Mohammad Tariq}
\authornote{Both authors contributed equally to this research.}
\email{ssmtariq@umich.edu}
\affiliation{%
  \institution{University of Michigan Dearborn}
  \city{Dearborn}
  \state{Michigan}
  \country{USA}
}

\author{Ali Al Zein}
\authornotemark[1]
\email{alielzei@umich.edu}
\affiliation{%
  \institution{University of Michigan Dearborn}
  \city{Dearborn}
  \state{Michigan}
  \country{USA}
}

\author{Soumya Sripad Vaidya}
\email{soumyasv@umich.edu}
\affiliation{%
  \institution{University of Michigan Dearborn}
  \city{Dearborn}
  \state{Michigan}
  \country{USA}
}

\author{Arati Khanolkar}
\email{aratik@umich.edu}
\affiliation{%
  \institution{University of Michigan Dearborn}
  \city{Dearborn}
  \state{Michigan}
  \country{USA}
}

\author{Probir Roy}
\email{probirr@umich.edu}
\affiliation{%
  \institution{University of Michigan Dearborn}
  \city{Dearborn}
  \state{Michigan}
  \country{USA}
}

\renewcommand{\shortauthors}{Tariq et al.}

\begin{abstract}
Serverless computing abstracts away server management, enabling automatic scaling and efficient resource utilization. However, cold-start latency remains a significant challenge, affecting end-to-end performance. Our preliminary study reveals that inefficient library initialization and usage are major contributors to this latency in Python-based serverless applications. We introduce \tool{}, a Python profiler that uses dynamic program analysis to identify inefficient library initializations. \tool{} collects library usage data through statistical sampling and call-path profiling, then generates a report to guide developers in addressing four types of inefficiency patterns. Systematic evaluations on 15 serverless applications demonstrate that \tool{} effectively identifies inefficiencies. \tool{} guided optimization results up to 2.26$\times$ speedup in cold-start execution time and 1.51$\times$ reduction in memory usage.
\end{abstract}




\received{20 February 2007}
\received[revised]{12 March 2009}
\received[accepted]{5 June 2009}

\maketitle




\section{Introduction}
\label{Intro}
The evolution of cloud computing has given rise to the increasingly popular concept of serverless computing. Serverless computing abstracts server management, enabling developers to build and run applications without the inconvenience of managing infrastructure. Serverless computing allows resources to scale according to application needs, providing fine-grained control that enhances resource utilization among cloud tenants. Additionally, it offers a more efficient pricing model, where users only pay for the resources their functions consume.

\begin{figure}
    \setlength{\belowcaptionskip}{-15pt}
    \includegraphics[width=\linewidth]{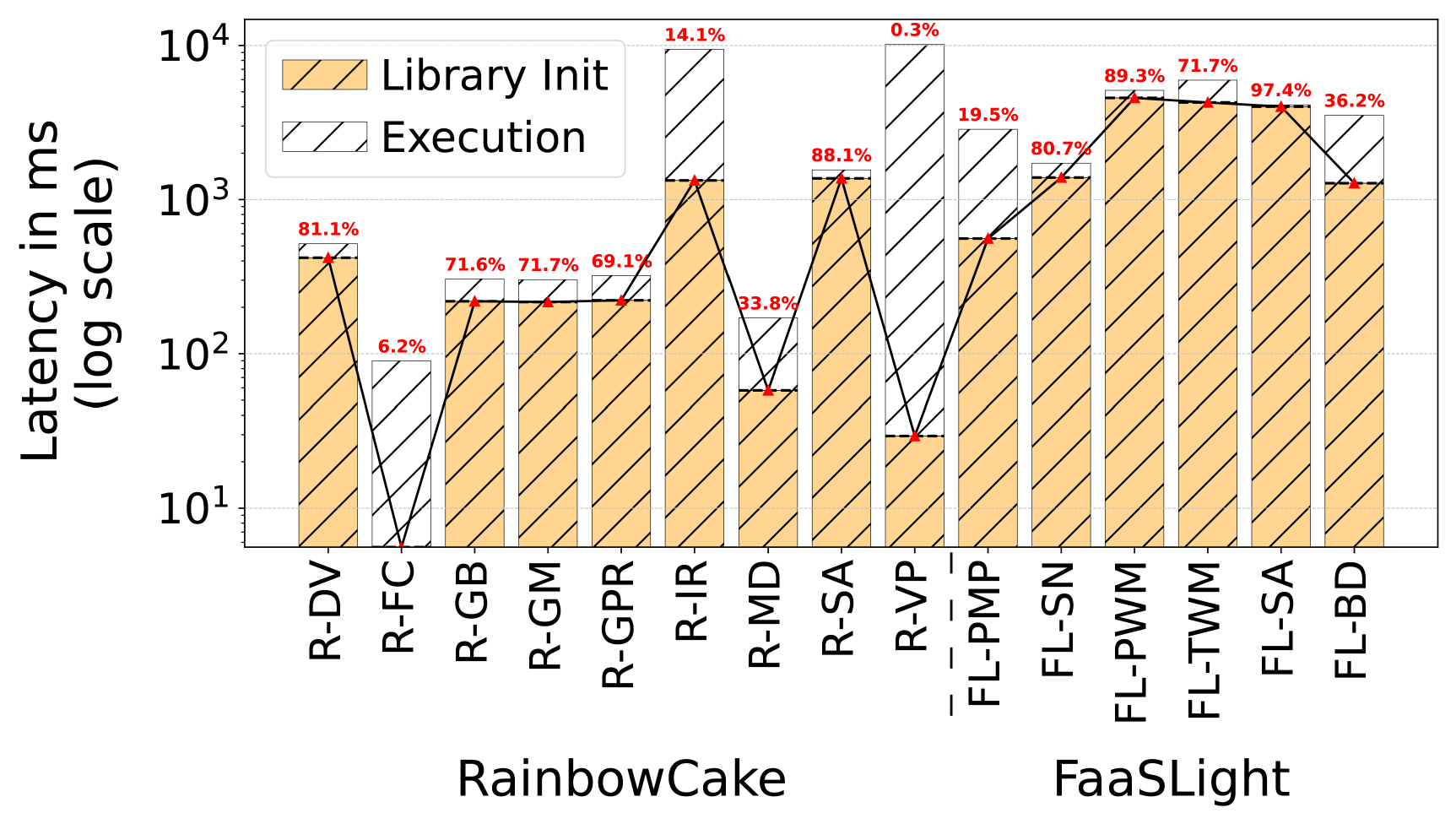}
    \caption{Ratio of library Initialization time to application execution time.}
    \label{fig:init_to_e2e}
\end{figure}

Despite its advantages, serverless computing faces several challenges. One significant issue is the ephemeral nature of serverless functions: once they stop being invoked and idle past a set \textit{keep-alive} period, their resources are reclaimed for other workloads. When the function is invoked again, it must be re-initialized, a process known as a \textbf{cold start}~\cite{shahrad2020serverless, wang2018peeking}. Because function invocations often occur in unpredictable patterns, frequent re-initialization negatively impacts response latency. Previously, research found response latency is directly related to end-user satisfaction and company revenues. In a study in 2006, Amazon reported a 100ms increase in latency cost them 1\% in sales~\cite{amazon_study}. Similar study at Google reported an extra 500ms in Google response time drops traffic by 20\%~\cite{google_study}. 

There have been significant efforts to minimize the impact of cold-start times in serverless computing. System research community has explored platform-level runtime optimizations, such as improving shared resource utilization~\cite{li2022help}, automatic memory deduplication~\cite{saxena2022memory}, caching~\cite{chen2023s}, efficient scheduling~\cite{pan2023sustainable}, and instance reuse~\cite{bhasi2021kraken, gunasekaran2020fifer, roy2022icebreaker, shahrad2020serverless}. Serverless cloud providers enable features to the developers for serverless runtime optimizations such as checkpointing~\cite{ao2022faasnap, du2020catalyzer, silva2020prebaking}, provisioned concurrency~\cite{provisionedConcurrencyAWS}, memory and compute resource allocation~\cite{optimisingServerlessForBBC, improveColdstartByIncreasingMemory}, keep-alive configurations~\cite{fuerst2021faascache, pan2022retention, roy2022icebreaker, shahrad2020serverless}, and dependency and execution Environment layering. Despite the efforts, cold-start is a significant challenge for latency-sensitive serverless applications.

In this paper, we address the issue of serverless cold-start latency from the perspective of code optimization, specifically focusing on applications written in Python. Our findings indicate that a significant source of cold-start latency in Python applications arises from library initialization. To quantify the impact of library initialization on overall execution time, we evaluated a collection of serverless Python applications drawn from existing literature~\cite{yu2024rainbowcake,liu2023faaslight}. Figure \ref{fig:init_to_e2e} presents the library initialization time, execution time, and their respective ratios. The results demonstrate that, for the majority of serverless applications, library initialization contributes to more than 70\% of the total execution time. Through a detailed manual investigation of these applications, we identified a considerable number of libraries or modules that are either unnecessary or could be avoided. However, pinpointing such optimization opportunities within the code is a challenging task.

Despite serverless functions often being small, typically comprising a few hundred lines of code, they frequently depend on external libraries for core functionality. These external libraries are often substantial, containing hundreds or thousands of lines of code, and are not originally designed with serverless environments in mind. Since, in long-running monolithic applications, the cost of library initialization is minimal, library developers often overlook actual code utilization. As a result, these libraries frequently perform inefficient module initialization. Furthermore, these libraries often depend on additional external libraries, creating a chain of dependencies that further increases library initialization time. Due to the complexity of these dependencies and the large code-bases of the libraries, manually identifying code optimization opportunities becomes challenging. 

Current static analysis techniques, such as FaaSLight~\cite{liu2023faaslight}, perform reachability analysis to determine whether a library is being invoked by a serverless entry function. However, these static analysis techniques have several limitations. First, they are unable to resolve dynamic contexts such as alias analysis, resulting in suboptimal solutions. Second, static analysis tools do not rely on initialization time measurements, which leads to a lack of prioritization in optimization efforts. Consequently, this may result in over-optimization, which negatively impacts the execution time.

To address this challenge, we propose \tool{}, a Python profiler that performs dynamic program analysis to identify inefficient library initialization in serverless applications written in Python. \tool{} performs statistical sampling and call-path profiling during serverless application invocations to collect data on library usage and their calling contexts. This technique accumulates samples and calling contexts from a large number of serverless application invocations to provide deep insights into library usage patterns. \tool{} is implemented as a Python module that application developers can attach while running their serverless applications. It does not require privileged access to the system, allowing it to be deployed seamlessly in production cloud Environments.

We perform an extensive evaluation to verify the effectiveness of \tool{} using three serverless application benchmarks and two real-world serverless applications. Guided by \tool{}, we identify 4 types of inefficient library usage patterns that significantly increase cold-start time. After eliminating these inefficiencies, we achieve up to 2.30$\times$ speedup in cold-start initialization time and 2.26$\times$ speedup in cold-start execution time. Additionally, the optimizations led to a 1.51$\times$ reduction in memory usage. 

\paragraph{\textbf{Paper Contributions}}
In summary, this paper makes the following  contributions:

\begin{itemize}
\item This paper is the first to conduct an in-depth study of inefficiency patterns in library initialization and their impact on serverless cold-start times.
\item We implement \tool{}, a profiler designed to identify inefficient library usage patterns in serverless applications. \tool{} does not require privileged access to the system or modifications to the Python runtime, making it suitable for adoption in serverless applications running in production cloud Environments.
\item Guided by \tool{}, we identified 4 inefficiency patterns in 15 applications. Optimizing these inefficiencies resulted in a significant speedup in the cold-start execution time of the serverless applications.
\end{itemize}

\begin{figure*}[ht]
    \centering
    \begin{tikzpicture}[scale=1, every node/.style={scale=1}]
        \draw[->] (0, 0) -- (17, 0) node[below] at (16.5, -0.1) {Time};

        \filldraw[fill=brown!10, draw=black, rounded corners] (0, 0) rectangle (1.5, 1);
        \node at (0.75, 0.5) {\small \begin{tabular}{c}Environment\\Init.\end{tabular}};
        
        \filldraw[fill=gray!10, draw=black, rounded corners] (1.5, 0) rectangle (3, 1);
        \node at (2.25, 0.5) {\small \begin{tabular}{c}Container\\Init.\end{tabular}};
        
        \filldraw[fill=red!10, draw=black, rounded corners] (3, 0) rectangle (4.5, 1);
        \node at (3.75, 0.5) {\small \begin{tabular}{c}Function\\Init.\end{tabular}};

        \filldraw[fill=white!20, draw=black, rounded corners] (4.5, 0) rectangle (6, 1);
        \node at (5.25, 0.5) {\small \begin{tabular}{c}Execution\\Time\end{tabular}};

        \filldraw[fill=white!20, draw=black, rounded corners] (6.5, 0) rectangle (8, 1);
        \node at (7.25, 0.5) {\small \begin{tabular}{c}Execution\\Time\end{tabular}};

        \draw[black, thick, <->] (6, -0.3) -- (9, -0.3);
        \node[above] at (8, -0.8) {\small \textit{Keep Alive}};

        \filldraw[fill=brown!10, draw=black, rounded corners] (10.5, 0) rectangle (12, 1);
        \node at (11.25, 0.5) {\small \begin{tabular}{c}Environment\\Init.\end{tabular}};
        
        \filldraw[fill=gray!10, draw=black, rounded corners] (12, 0) rectangle (13.5, 1);
        \node at (12.75, 0.5) {\small \begin{tabular}{c}Container\\Init.\end{tabular}};
        
        \filldraw[fill=red!10, draw=black, rounded corners] (13.5, 0) rectangle (15, 1);
        \node at (14.25, 0.5) {\small \begin{tabular}{c}Function\\Init.\end{tabular}};

        \filldraw[fill=white!20, draw=black, rounded corners] (15, 0) rectangle (16.5, 1);
        \node at (15.75, 0.5) {\small \begin{tabular}{c}Execution\\Time\end{tabular}};

        \node[above] at (2, 1.2) {\small \textit{Cold Start}};
        \node[above] at (7.25, 1.2) {\small \textit{Warm Start}};
        \node[above] at (13, 1.2) {\small \textit{Cold Start}};

        \draw[thick, dashed] (0, 1) -- (0, 1.5) node[above] {\small \begin{tabular}{c}Function\\Invocation\end{tabular}};
        \draw[thick, dashed] (4.5, 1) -- (4.5, 1.5) node[above] {\small \begin{tabular}{c}End Cold\\Start\end{tabular}};
        \draw[thick, dashed] (6.5, 1) -- (6.5, 1.5) node[above] {\small \begin{tabular}{c}Second\\Invocation\end{tabular}};
        \draw[thick, dashed] (9, 0) -- (9, 1.5) node[above] {\small \begin{tabular}{c}\textit{Keep Alive}\\End\end{tabular}};
        \draw[thick, dashed] (10.5, 1) -- (10.5, 1.5) node[above] {\small \begin{tabular}{c}Third\\Invocation\end{tabular}};
        \draw[thick, dashed] (15, 1) -- (15, 1.5) node[above] {\small \begin{tabular}{c}End Cold\\ Start\end{tabular}};
        
        \draw[black, thick, <->] (0, 1.2) -- (4.5, 1.2);

        \draw[black, thick, <->] (10.5, 1.2) -- (15, 1.2);

        \draw[thick, dashed] (6, 0) -- (6, -0.3);
        \draw[thick, dashed] (9, 0) -- (9, -0.3);

    \end{tikzpicture}
    \caption{Timeline of serverless function lifecycle events, illustrating stages from environment initialization to function execution, including cold start, warm start, and keep-alive periods.}
    \label{fig:serverless_lifecycle}
\end{figure*}
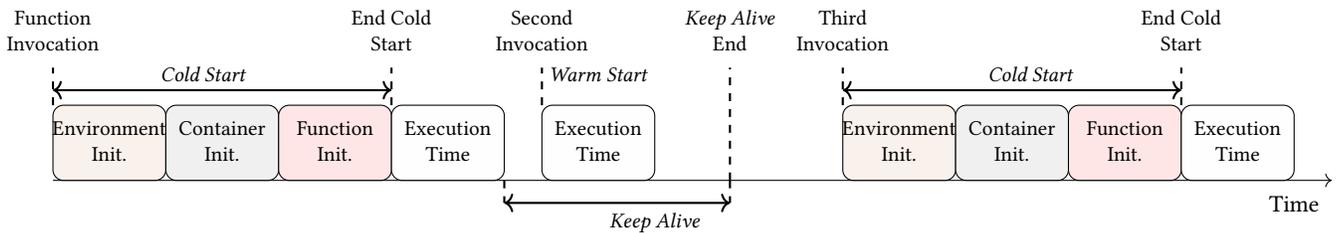

\section{Background and Related work}
\label{background}



\subsection{Serverless Cold-start}

Figure~\ref{fig:serverless_lifecycle} illustrates the lifecycle of a serverless function. The initial function invocation triggers a cold start, which involves a three-phase initialization process: environment bootstrapping, container initialization, and function setup. Once the function is initialized, it proceeds to execution. Typically, the function container remains active for a predefined \textit{keep-alive} period. Subsequent invocations within this \textit{keep-alive} period result in a warm start, allowing the function to execute without undergoing the re-initialization phases. Upon expiration of the keep-alive period, the container is reclaimed, and any subsequent invocations will trigger a cold start, necessitating the three-phase re-initialization process. The cold start latency is influenced by these three phases:

\begin{itemize}
    \item \textbf{Environment initialization} involves setting up the infrastructure environment including setting up network connections, loading configuration files, and starting background processes to provide functionalities such as logging, monitoring, and security enhancements.
    \item \textbf{Container initialization} entails setting up the runtime environment needed to execute the serverless function. This includes loading the runtime binaries and libraries necessary for the function's execution. The runtime could be based on various programming languages (e.g., Python, Node.js, Java), and this phase ensures that all required dependencies and resources for the runtime are available and configured correctly.
    \item \textbf{Function initialization} includes loading the actual function code and its dependent libraries into memory. This phase involves running any static initializers and setting up global variables as defined in the function code. It ensures that the function is fully prepared to handle incoming events or requests by completing any necessary setup tasks within the function itself. \textit{This paper addresses the library initialization occurring in this third phase}.
\end{itemize}

\subsection{Mitigation of Serverless Cold-start}
\label{mitigating_coldstart}
In this section, we briefly introduce related works from the system and software engineering research community that aim to mitigate cold start times.
\paragraph{\textbf{Platform-Level Runtime Optimizations:}}
Several techniques have been proposed to enhance infrastructure efficiency and mitigate cold start latency through optimized resource allocation and scheduling of serverless functions. These methodologies encompass shared resource utilization~\cite{li2022help}, automatic memory deduplication~\cite{saxena2022memory}, function caching~\cite{chen2023s}, compression~\cite{basu2024codecrunch}, advanced scheduling algorithms~\cite{pan2023sustainable}, and the reuse of pre-warmed instances~\cite{bhasi2021kraken, gunasekaran2020fifer, roy2022icebreaker, shahrad2020serverless}.
\paragraph{\textbf{User-Directed Serverless Runtime Optimizations:}} User-directed optimizations involve configuring serverless runtime policies to reduce cold start times. Techniques include checkpointing~\cite{ao2022faasnap, du2020catalyzer, silva2020prebaking} to save function state, provisioned concurrency~\cite{provisionedConcurrencyAWS} to keep instances warm, adjusting memory~\cite{improveColdstartByIncreasingMemory} and compute resources~\cite{optimisingServerlessForBBC} to optimize performance, keep-alive~\cite{fuerst2021faascache, pan2022retention, roy2022icebreaker, shahrad2020serverless} configurations to prevent premature termination, and layering dependencies~\cite{yu2024rainbowcake} to reduce loading overhead by caching and updating them independently.
\paragraph{\textbf{Code-level optimizations:}} 
These involve modifying serverless application code to reduce initialization time and enhance performance. Techniques include function fusion~\cite{lee2021mitigating}, which combines multiple serverless functions to minimize initialization overhead, function decomposition~\cite{kalia2021mono2micro, nitin2022cargo, abgaz2023decomposition}, which breaks down complex serverless functions into smaller, more manageable units, and serverless function compression~\cite{liu2023faaslight}, which reduces code size and dependencies to decrease load times. 

The work most closely related to \tool{} is FaaSLight~\cite{liu2023faaslight}, which utilizes static analysis to identify opportunities and refactor code for serverless function compression. However, static analysis techniques are not guided by performance metrics and, therefore, cannot prioritize optimization opportunities. Additionally, static analysis tools are unable to observe dynamic library usage patterns. They also often struggle with analyzing complex code, such as in alias analysis. These limitations can lead to suboptimal results in performance optimization. In contrast, \tool{} employs dynamic program analysis with a measurement-based approach to identify optimization opportunities. Unlike FaaSLight, \tool{} does not refactor code but instead guides the developer by highlighting potential optimization opportunities, allowing the developer to select the appropriate optimization strategy for the specific scenario. In section~\ref{eval}, we conduct a comparative performance evaluation between \texttt{FaaSLight} and \tool{}.

\subsection{Python Library Structure and it's Initialization}
\label{python_structure}
\paragraph{\textbf{Library, Package and Module}}
In Python, a \textbf{library} is a collection of related modules and packages that provide a wide range of functionalities aimed at specific tasks or domains. A \textbf{module} is a single file containing Python code—functions, classes, or variables—that can be imported and used in other Python programs to promote code reuse and organization. A \textbf{package}, on the other hand, is a collection of related modules organized within a directory. Packages provide a hierarchical structure to the code, facilitating better organization and modularization.

Packages structure their constituent modules using a hierarchical directory format. The root package directory includes an \texttt{\_\_init\_\_.py} file and multiple Python files (.py), each representing a module. Subdirectories within the root directory serve as subpackages, each containing their own \texttt{\_\_init\_\_.py} files and additional modules. This structured approach allows independent maintenance of different components while ensuring they collectively form a cohesive package.

\paragraph{\textbf{Library initialization time}}
When a package within a library is imported, Python executes the \texttt{\_\_init\_\_.py} files and any other code in the imported modules. This process sets up necessary variables, configurations, and class definitions. The total initialization time for a library is the sum of the initialization times of its individual modules and packages. Each package's \texttt{\_\_init\_\_.py} file may contain setup code that runs during import, including importing other modules, initializing variables, and executing startup routines. Thus, the library's initialization time is the accumulated initialization times of its packages and modules.





\subsection{Python Profilers}
Various Python performance profiling tools have been developed over the years to address specific aspects of performance analysis. At a high level, we divide approaches into instrumentation and sampling-based profilers. The instrumentation-based profilers such as Python's Built-in \textit{cProfile}~\cite{cprofile} and \textit{Profile}~\cite{profile}, trace Python applications at the function or line-level granularity. Due to tracing, these instrumentation-based profilers incur significant overhead and inaccuracy in performance measurements~\cite{288540}. The sampling-based profilers, such as \textit{py-spy}~\cite{py-spy}, \textit{Austin}~\cite{austin}, \textit{Scalene}~\cite{288540}, \textit{Pprofile}~\cite{Pprofile}, and \textit{Pieprof}~\cite{tan2021toward} incur less overhead and provide performance insights at relatively higher accuracy. However, none of these profilers identify inefficient library usage in Python code. In contrast, \tool{} is designed to improve cold-start times of serverless applications by identifying inefficient library initialization.
 
\section{Characterizing Cold-start inefficiency}
\label{motive}
This section outlines various inefficient library usage patterns identified by \tool{} during the evaluation of 15 serverless applications. To the best of our knowledge, this is the first study to characterize these inefficiencies, providing insights into how they contribute to increased cold-start times. Additionally, we offer concrete examples of each identified inefficient categories, which was instrumental in the development of \tool{}.

\subsection{Cold-start Inefficiency Categories}
\label{ineff_category}
In this section, we examine four inefficiency patterns in library usage within serverless applications. 

\begin{table}[!htbp]
    \centering
    \small
    \begin{tabular}{>{\raggedright\arraybackslash}p{7.5cm}}
        \hline
        \rowcolor{gray!30}\textbf{File: igraph/clustering.py, Lines 11-13} \\
        \hline
        \texttt{from igraph.drawing.colors import ...} \\
        \texttt{from igraph.drawing.cairo.dendrogram import ...} \\
        \texttt{from igraph.drawing.matplotlib.dendrogram import ...} \\
        \hline
        \rowcolor{gray!30}\textbf{Call Path} \\
        \hline
        handler.py:2 \\
        \quad-> igraph/\_\_init\_\_.py:104 \\
        \quad\quad-> igraph/community.py:2 \\
        \quad\quad\quad-> igraph/clustering.py:<11-13> \\
        \hline
    \end{tabular}
    \caption{C1 - Importing unused libraries in \textit{graph\_bfs}.}
    \label{tab:motiv_example_unused_lib}
\end{table}

\begin{table}[!htbp]
    \centering
    \small
    \begin{tabular}{>{\raggedright\arraybackslash}p{7.5cm}}
        \hline
        \rowcolor{gray!30}\textbf{File: cve\_bin\_tool/validator.py, Line 11} \\
        \hline
        \texttt{import xmlschema} \\
        \hline
        \rowcolor{gray!30}\textbf{Call Path} \\
        \hline
        handler.py:11 \\
        \quad-> cve\_bin\_tool/cli.py:71 \\
        \quad\quad-> cve\_bin\_tool/sbom\_detection.py:8 \\
        \quad\quad\quad-> cve\_bin\_tool/validator.py:11 \\
        \hline
    \end{tabular}
    \caption{C2 - Import rarely used libraries by default in CVE Binary Tool.}
    \label{tab:motivExampleRarelyUsedLib}
\end{table}

\begin{table}[!htbp]
    \centering
    \small
    \begin{tabular}{>{\raggedright\arraybackslash}p{7.5cm}}
        \hline
        \rowcolor{gray!30}\textbf{Pull request: Replace pkg\_resources \#358} \\
        \hline
            \begin{tikzpicture}
            \node[rectangle, draw=none, text width=7.5cm, inner sep=5pt] (m) {
                \large ``Just importing pkg\_resources becomes slower and slower the more packages are installed. Even on fast machines with SSDs this can add a second or more to the startup time.''
            };
        \end{tikzpicture} \\
        \hline
        \rowcolor{gray!30}\textbf{File: docs/conf.py, Lines 16} \\
        \hline
        \rowcolor{red!50} \texttt{- import pkg\_resources} \\
        \rowcolor{green!50} \texttt{+ from importlib.metadata import version} \\
        \hline
    \end{tabular}
    \caption{C3 - Pull request showing library misuse in \textit{Chameleon}.}
    \label{tab:motiv_example_lib_misuse}
\end{table}

\begin{table}[!htbp]
    \centering
    \small
    \begin{tabular}{>{\raggedright\arraybackslash}p{7.5cm}}
        \hline
        \rowcolor{gray!30}\textbf{File: ocrmypdf/\_logging.py, Lines 11} \\
        \hline
        \rowcolor{red!50}\texttt{- from rich.console import Console} \\
        \rowcolor{red!50}\texttt{- from rich.logging import RichHandler} \\
        \hline
        \rowcolor{gray!30}\textbf{Call Path } \\
        \hline
        apply-ocr-to-s3-object.py:6 \\
        \quad-> ocrmypdf/\_\_init\_\_.py:14 \\
        \quad\quad-> ocrmypdf/api.py:18 \\
        \quad\quad\quad-> ocrmypdf/logging.py:11 \\
        \hline
    \end{tabular}
    \caption{C4 - Removing library without impacting core functionality in \textit{OCRmyPDF}}
    \label{tab:motiv_example_replacable_lib}
\end{table}
\paragraph{\textbf{C1 - Importing Unused Libraries}}
Table~\ref{tab:motiv_example_unused_lib} presents a code snippet from the RainbowCake-\textit{graph\_bfs} serverless application and its dependency on the \textit{igraph} library. The \textit{graph\_bfs} application executes a breadth-first search on a generated graph using the \textit{igraph} library. \textit{igraph} is a comprehensive toolset for graph analysis, including robust graph visualization capabilities. When imported by the \textit{graph\_bfs} application, \textit{igraph} initializes many features by default, including its visualization tools. However, the \textit{graph\_bfs} application only utilizes \textit{igraph} for graph traversal, making the initialization of the visualization capabilities unnecessary. Our experiments indicate that \textit{igraph}'s visualization tool contributes to a 37\% overhead in initialization time for the \textit{graph\_bfs} application. By customizing the \textit{igraph} library package to disable the initialization of the visualization tool and other non-essential components, we achieved a 1.65$\times$ improvement in the library's initialization time for the application.
\paragraph{\textbf{C2 - Default Import of Rarely Used Libraries}}
The CVE Binary Tool identifies known vulnerabilities in software packages. It integrates into continuous integration systems for regular scanning and early warnings. We deploy it as a serverless function. It scans component lists in various formats, including .csv files, Linux package lists, language-specific scanners, and Software Bill of Materials (SBOM) formats.

Although SBOM files are rarely found in our test dataset, the CVE Binary Tool loads the SBOM validator and its dependent libraries by default.  One such library, \textit{xmlschema}, has 8\% initialization overhead. Table~\ref{tab:motivExampleRarelyUsedLib} shows the call path of the library call. Since SBOM files and their validation are infrequent, we can reduce this overhead by implementing lazy loading. This optimization will prevent unnecessary library loading in most cases of serverless CVE Binary Tool invocation, though a few invocations that require SBOM validation will still need to load the dependent libraries.

\paragraph{\textbf{C3 - Library Misuse}}
\textit{Chameleon}, a Python HTML template engine, uses \textit{pkg\_resources} from the setuptools package to manage and locate resources within Python packages. While \textit{pkg\_resources} incurs significant overhead, \textit{importlib\_resources} is a modern alternative for resource management in Python. It is part of the \textit{importlib} module, included in the Python standard library starting from Python 3.7. \textit{importlib\_resources} is designed to be more efficient and lightweight compared to \textit{pkg\_resources}. Despite the overhead, Chameleon previously relied on \textit{pkg\_resources}. However, as shown in table~\ref{tab:motiv_example_lib_misuse}, the developers addressed the API misuse in version 4.1.0 and onwards by replacing it with \textit{importlib}.

\paragraph{\textbf{C4 - Avoidable Library Usage Without Impacting Core Functionality}}
Table~\ref{tab:motiv_example_replacable_lib} provides an example of avoidable library usage. In this example, \textit{OCRmyPDF}, an Optical Character Recognition (OCR) tool, relies on the \textit{rich} library. The \textit{rich} library is a Python package that offers advanced formatting and styling capabilities for terminal output. It can render text with different styles, colors, tables, progress bars, and even markdown and syntax-highlighted code. OCRmyPDF uses the \textit{rich} library to display progress bar information and other logging tools related to PDF processing. However, the library incurs 13\% overhead in library initialization. This usage is avoidable without impacting OCRmyPDF's core functionality.
    

\subsection{Design Goal of \tool{}}
Based on the four inefficiency patterns, we make the following observations that are instrumental in designing \tool{}:

\begin{itemize}
    
    \item Due to the potentially large number of dependent libraries, it is essential to first identify which libraries contribute to \textbf{significant overhead} during initialization, thereby impacting the performance of serverless cold-starts.
    
    \item Since the static analysis has limited context and cannot accurately capture the dynamic behavior of an application, it is essential to monitor the application's library usage patterns \textbf{at runtime} to determine whether a loaded library is being utilized.
    
    \item Considering that library usage depends on application and workload characteristics, it is important to monitor library usage over a large number of serverless function invocations. This will enable the observation of the \textbf{frequency of library usage} and thus identify rarely used libraries.
    
    \item Since libraries can be invoked from different contexts, it is important to capture the \textbf{calling context} of library usage to pinpoint the source of inefficiencies.

    \item Because serverless applications are latency-sensitive, the \textbf{monitoring overhead} per serverless function invocation must be kept minimal to be useful and practical.
\end{itemize}

\section{\tool{} Design and Implementation}
\subsection{Overview}

\tool{} employs a systematic methodology to identify inefficient library usage patterns in applications that result in notable performance overhead. Initially, (1) to reduce the effort of the analysis, \tool{} calculates the ratio between the library initialization time and the execution time of serverless functions. If this ratio falls below a predefined threshold, $T$, the library is considered to have an insignificant impact on the cold start. Applications exhibiting significant library initialization times are subjected to more detailed dynamic program analysis. In this study, the threshold is set at 10\%.

One approach to dynamically monitor applications at runtime is instrumentation. However, prior studies have shown that code or binary instrumentation incurs significant runtime overhead, making the adoption of monitoring tools in production workloads impractical.  As an alternative, \tool{} leverages (2) statistical sampling to profile serverless function invocations. During statistical sampling, \tool{} collects call-paths at each sampling point to construct a comprehensive calling context. Then, \tool{} (3) attributes these call-paths to their respective libraries and calculates a library utilization metric. This utilization metric, along with detailed call-path profiles, helps to pinpoint library usage patterns and their inefficiencies. The following sections provide an in-depth discussion of these three components.
 


\begin{figure}
    \setlength{\belowcaptionskip}{-15pt}
    \scalebox{0.5}{\input{figures/arch_v2}}
    \caption{\tool{} architecture overview.}
    \label{fig:architecture}
\end{figure}

\subsection{Hierarchical Breakdown of Initialization Overhead}
\label{init_time_measurement}
\tool{} employs a principled approach to identify libraries and packages that incur significant initialization overhead, thereby impacting cold-start times. Central to \tool{}'s methodology is a hierarchical analysis policy, which systematically decomposes the initialization overhead from the application level down to individual libraries, and further into their constituent packages and sub-packages. This thorough approach provides a comprehensive and detailed view of the specific areas where optimization efforts should be concentrated, facilitating targeted improvements in cold-start performance.

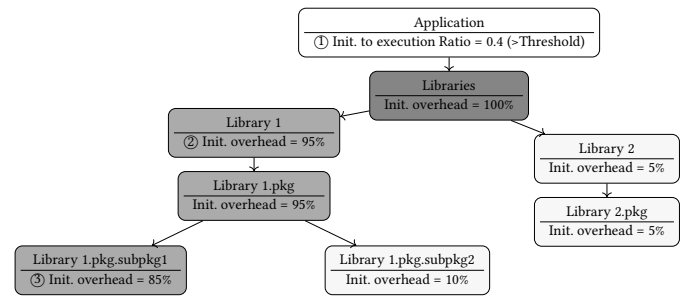
\begin{figure}[h!]
\centering
\resizebox{0.5\textwidth}{!}{%
\begin{tikzpicture}[node distance=2cm, auto]

\node (app) [draw, rectangle, rounded corners, text centered, font=\small] {\begin{tabular}{c}
    Application \\
    \hline
    \textcircled{\small{1}} Init. to execution Ratio = 0.4 (>Threshold)
  \end{tabular}};
\node (libs) [draw, rectangle, rounded corners, below of=app, text centered, node distance=1.2cm, font=\small, fill=gray!80] {
  \begin{tabular}{c}
    Libraries \\
    \hline
    Init. overhead = 100\%
  \end{tabular}
};
\node (lib1) [draw, rectangle, rounded corners, below left of=libs, xshift=-3cm, text centered, node distance=1cm, font=\small, fill=gray!55] {
  \begin{tabular}{c}
    Library 1 \\
    \hline
    \textcircled{\small{2}} Init. overhead = 95\%
  \end{tabular}
};
\node (lib2) [draw, rectangle, rounded corners, below of=libs, xshift=3cm, text centered, node distance=1.2cm, font=\small, fill=gray!5] {
  \begin{tabular}{c}
    Library 2 \\
    \hline
    Init. overhead = 5\%
  \end{tabular}
};
\node (pkg1) [draw, rectangle, rounded corners, below of=lib1, text centered, node distance=1.2cm, font=\small, fill=gray!55] {
  \begin{tabular}{c}
    Library 1.pkg \\
    \hline
    Init. overhead = 95\%
  \end{tabular}
};
\node (pkg2) [draw, rectangle, rounded corners, below of=lib2, text centered, node distance=1.2cm, font=\small, fill=gray!5] {
  \begin{tabular}{c}
    Library 2.pkg \\
    \hline
    Init. overhead = 5\%
  \end{tabular}
};
\node (subpkg1) [draw, rectangle, rounded corners, below left of=pkg1, xshift=-1.5cm, text centered, node distance=2cm, font=\small, fill=gray!55] {
  \begin{tabular}{c}
    Library 1.pkg.subpkg1 \\
    \hline
    \textcircled{\small{3}} Init. overhead = 85\%
  \end{tabular}
};
\node (subpkg2) [draw, rectangle, rounded corners, below right of=pkg1, xshift=1.5cm, text centered, node distance=2cm, font=\small, fill=gray!5] {
  \begin{tabular}{c}
    Library 1.pkg.subpkg2 \\
    \hline
    Init. overhead = 10\%
  \end{tabular}
};

\draw[->] (app) -- (libs);
\draw[->] (libs) -- (lib1);
\draw[->] (libs) -- (lib2);
\draw[->] (lib1) -- (pkg1);
\draw[->] (lib2) -- (pkg2);
\draw[->] (pkg1) -- (subpkg1);
\draw[->] (pkg1) -- (subpkg2);

\end{tikzpicture}
}
\caption{Hierarchical analysis policy with initialization overhead. Darker nodes represent higher overhead leading to increased cold-start time.}
\label{fig:hierarchical_analysis}
\end{figure}

Figure~\ref{fig:hierarchical_analysis} depicts the hierarchical analysis with an example. For this hierarchical analysis, \tool{} follows these steps: \textcircled{\small{1}} it measures the ratio of the total initialization time of all libraries to the application execution time, which highlights the impact of initialization on cold-start. \textcircled{\small{2}} for applications experiencing significant cold-start issues based on a high ratio value, \tool{} considers the initialization time for each library to pinpoint the main contributors to the overhead. \textcircled{\small{3}} by subdividing each library's initialization time into packages and sub-packages, \tool{} further ranks them by their impact, enabling targeted optimizations.

\paragraph{\textbf{Measuring initialization overhead}}

Initially, \tool{} measures the import time of individual modules at runtime. By leveraging the Python package structure discussed in section~\ref{python_structure}, \tool{} calculates the \textbf{(1) initialization time for packages and sub-packages} and the \textbf{(2) individual library initialization time} by aggregating the import times of individual modules. Finally, \tool{} calculates the \textbf{(3) total initialization overhead} by accumulating the initialization times of all the loaded libraries.

If the initialization time of the $j$-th module in a package or sub-package is denoted by $T_{\text{j}}$ and the total number of imported modules is $M$, then the initialization time $T_{\text{package}}$ is:
\begin{equation}
T_{\text{package}} = \sum_{j=1}^{M} T_j
\end{equation}

To calculate the individual library initialization time, \tool{} sums the initialization times of all modules within the library. Given $N$ is the total number of imported modules in the library and $T_{\text{i}}$ is the initialization time of the $i$-th module, the library initialization time $T_{\text{library}}$ is:
\begin{equation}
T_{\text{library}} = \sum_{i=1}^{N} T_i
\end{equation}

Finally, the accumulated initialization time of all loaded libraries used in the application is calculated. If there are $L$ libraries used, and the initialization time of the $k$-th library is denoted by $T_{\text{library}_k}$, then the accumulated initialization time $T_{\text{total\_initialization}}$ is:
\begin{equation}
T_{\text{total\_initialization}} = \sum_{k=1}^{L} T_{\text{library}_k}
\end{equation}

\subsection{Sampling-Based Call-path Profiling}
To monitor the library usage of a serverless application, we employ sampling-based call-path profiling. \tool{} sets up a timer with a configurable sampling frequency and registers a signal handler. When the timer expires, it triggers the signal handler, enabling \tool{} to capture the application's current state at that moment. \tool{} gathers the following data in the signal handler: (1) the source code line number, (2) the function name, (3) file path and (4) the call path leading to the function call. At the implementation level, \tool{} utilizes Python's native \texttt{traceback} module to extract the frames from the Python call stack and construct the call path.

\paragraph{\textbf{Calling context tree (CCT)}}
\tool{} accumulates the call paths to construct a Calling Context Tree (CCT)~\cite{ammons1997exploiting}, a data structure that captures the function calling context, as shown in figure~\ref{fig:architecture}. In the CCT, each node represents a function call, and the edges represent the caller-callee relationship. The root of the tree represents the entry point of the serverless function. A CCT captures the workload context, distinguishing between different calling contexts of the same function. If a function is called from various parts of the code, each call will have its own unique path in the CCT, preserving the calling context. 

\tool{} further augments the CCT with the sample count of each function call to measure the frequency of function executions under the observed workload. The calling context frequency allows identification of sources of inefficient invocations and guides developers in optimizing their code.

\subsection{Associating the Call-paths to Libraries}
\tool{} enhances the call paths by associating them with the loaded library modules. This is achieved by labeling the nodes of the CCT with the corresponding library module names. Python packages are organized in a hierarchical structure using directories, with modules and sub-packages arranged within these directories. To accurately identify the package and module names for each library, \tool{} leverages the file paths associated with each node in the CCT. This allows \tool{} to map each function call to its respective library module, providing a clear understanding of how external libraries are utilized within the application.

\paragraph{\textbf{Measuring the Library Utilization}} Finally, \tool{} calculates a utilization metric for each library. This metric is defined as:

\begin{equation}
U(L) = \frac{\sum_{f \in L} S(f)}{\sum_{f \in F} S(f)}
\end{equation}

where \( U(L) \) is the utilization of library \( L \), \( S(f) \) is the sample count for function \( f \), \( f \in L \) represents all functions in library \( L \), and \( f \in F \) represents all functions in the application. This metric provides a proportional measure of how frequently functions from a specific library are executed relative to the entire application, helping identify frequent and in-frequent library usage and potential areas for optimization.

\subsection{Detecting inefficient library usage}

\tool{} leverages both package initialization time (section~\ref{init_time_measurement}) and the library utilization metric to guide the detection of inefficient library usage.

\paragraph{Detecting unused library imports}
\tool{} analysis ranks the libraries in descending order based on their initialization latency. If any library incurs significant overhead without any usage samples, it suggests the library is unused and should be considered for optimization to improve cold-start time.

\paragraph{Detecting rarely-used library imports}
In a similar manner, \tool{} identifies libraries that contribute to significant initialization overhead by ranking them based on their initialization latency. If the utilization metric shows that these libraries are used infrequently, there is a potential benefit to lazy loading them. This approach can help remove the library initialization from the hot path, improving overall performance by delaying the loading of these libraries until they are actually needed in rare call paths. 

\paragraph{Detecting libraries with significant overhead}
\tool{} reports libraries that cause significant initialization overhead along with their calling context. This information helps developers pinpoint where to investigate further to justify the library usage. By doing so, it reveals opportunities to address library misuse and eliminate unnecessary libraries, as discussed in section~\ref{ineff_category}.  

\subsection{Accuracy Analysis}
In this section, we perform an accuracy analysis of \tool{}. To prove that a sampling-based statistical approach can correctly identify the usage frequency of different libraries within an application, including those that are rarely used, we utilize principles from probability theory and statistical inference.

Let \( N \) be the total number of independent and identically distributed (IID) samples collected over many serverless invocations. For simplicity, lets assume, each sample identifies one library that is used within the application. We denote \( p_i \) as the true usage frequency of library \( i \), which is the true probability that library \( i \) is used within the application. The estimated usage frequency, \( \hat{p_i} \), is the frequency calculated from the samples.

By the Law of Large Numbers (LLN), the sample mean \( \hat{p_i} \) converges to the true mean \( p_i \) as \( N \rightarrow \infty \). For each library \( i \), let \( X_i \) be a random variable indicating its usage in the application (1 if used, 0 if not). The IID samples imply that \( X_i \) are Bernoulli random variables with parameter \( p_i \). The sample mean \( \hat{p_i} \) is given by:

\[
\hat{p_i} = \frac{1}{N} \sum_{j=1}^{N} X_{ij}
\]




The Central Limit Theorem (CLT) provides further insight into the distribution of \( \hat{p_i} \). For large \( N \), the distribution of \( \hat{p_i} \) can be approximated by a normal distribution:

\[
\hat{p_i} \sim \mathcal{N}\left(p_i, \frac{p_i (1 - p_i)}{N}\right)
\]

This normal approximation allows us to construct confidence intervals for \( p_i \). To estimate \( p_i \) with high confidence, we construct a confidence interval around \( \hat{p_i} \). For a 95\% confidence interval, we use the standard normal critical value (approximately 1.96):

\[
\hat{p_i} \pm 1.96 \sqrt{\frac{\hat{p_i} (1 - \hat{p_i})}{N}}
\]

As \( N \) increases, the width of the confidence interval decreases, increasing our confidence in the estimate \( \hat{p_i} \).

For libraries that are rarely used, \( p_i \) is small. Despite this, as long as \( N \) is sufficiently large, \( \hat{p_i} \) will still be a reliable estimate due to the properties of Bernoulli trials and the LLN. If \( p_i \) is very small, the variance \( \frac{p_i (1 - p_i)}{N} \) is also small, provided \( N \) is large enough. This ensures that the confidence interval remains tight even for rarely used libraries.

In conclusion, by collecting a sufficiently large number of IID samples (\( N \)), we can estimate the usage frequency (\( p_i \)) of each library with high confidence. The convergence properties of the sample mean, along with the construction of confidence intervals, ensure that even for libraries that are rarely used, the frequency can be accurately detected. This statistical sampling approach is robust and forms the basis of many statistical sampling-based profiling tools.

\subsection{Implementation}
\tool{} is developed as a Python module. Application developers import \tool{} similarly to other libraries to monitor library usage patterns. At the implementation level, it performs call-path sampling and monitors library initialization time during serverless function execution. \tool{} aggregates sampled data from numerous invocations and stores it in cloud storage services such as AWS DynamoDB or S3. Subsequently, \tool{} conducts a post-mortem analysis on the profiled data to construct a calling context tree and calculate utilization metrics. An online result visualizer is also implemented within \tool{} to assist developers with the analysis results. This visualizer is hosted as a serverless application. It enables developers to sort libraries by initialization overhead and utilization. For a more detailed analysis of library usage patterns, the visualizer also presents the calling context tree of the library. The \tool{} and its visualizer are included in the replication package. 
\section{Evaluation}
\label{eval}
In this section we evaluate the utility, impact, and overhead incurred by \tool{}. We assess \tool{} focusing on answering the following questions:
\begin{itemize}
    \item \textbf{Q1 (Detection):} Does \tool{} effectively identify and report inefficient Python library usage? 
    \item \textbf{Q2 (Speedup):} How much performance improvement can be expected through \tool{}-guided optimization?
    \item \textbf{Q3 (Overhead):} How much overhead \tool{} imposes in its default setting? 
\end{itemize}

\paragraph{\textbf{Experimental setup}}
We evaluate \tool{} on the AWS Lambda service. All the serverless applications were deployed as AWS Lambda functions, using the Python 3.9 runtime, with a default memory size of 1024MB, a storage size of 1024MB, and a timeout of 60 seconds. Given the diverse use cases of the multiple applications evaluated, some applications also utilize other AWS-managed services such as S3 and Amazon Elastic Container Registry (ECR). Each application's dependencies and source code are packaged into a zip file, and functions are created by transmitting the zip from an S3 bucket. Applications that use large Python libraries, such as \texttt{Pytorch}, \texttt{torchvision}, and \texttt{cve-bin-tool}, are deployed using Docker container images through AWS ECR with different memory and timeout configurations. Function URLs are created for each AWS Lambda function, and the appropriate roles are configured

To assess the effectiveness of \tool{}, we examined 20 serverless applications. This includes two real-world applications: CVE-bin-tool and OCRmyPDF. The remaining applications were selected from three popular benchmarks: RainbowCake~\cite{yu2024rainbowcake}, Faaslight~\cite{liu2023faaslight}, and FaaSWorkbench~\cite{kim2019functionbench}.


\paragraph{\textbf{Evaluation methodology}}

To evaluate \tool{}, we examine each application across three execution phases. In each phase, we request the corresponding Lambda functions to record initialization latency, execution latency, and peak memory usage from the AWS CloudWatch logs for 500 cold starts and report results by averaging the outcomes from five iterative executions of each phase.

First, we execute the unmodified applications and record the aforementioned metrics. We then measure the initialization overhead of each application by calculating the ratio of initialization latency to execution latency. Applications with less than 10\% initialization overhead are considered to have insignificant initialization time and are discarded from further evaluation. Next, to identify inefficiencies, we run the application along with \tool{} and collect the inefficiency reports. Guided by the \tool{} inefficiency reports, we further optimize the applications. In the final phase, we deploy the optimized applications, send requests, and collect the same metrics.

\begin{table*}
\fontsize{7.5pt}{8.5pt}\selectfont
\begin{tabular}{@{}ccccccccc@{}}
\hline
\multicolumn{3}{c}{\textbf{Program Information}} & \multicolumn{1}{c}{\textbf{Inefficiency}} & \multicolumn{4}{c}{\textbf{Speedup}} \\ \hline
\textbf{Applications} & \textbf{Library} & \textbf{Type} & \textbf{Category} & \textbf{\begin{tabular}[c]{@{}c@{}}Initialization\\Speedup \\(times)\end{tabular}} & \textbf{\begin{tabular}[c]{@{}c@{}}Execution \\ Speedup \\(times)\end{tabular}} & \textbf{\begin{tabular}[c]{@{}c@{}}99\textsuperscript{th} Percentile \\ Initialization \\Speedup\end{tabular}} & \textbf{\begin{tabular}[c]{@{}c@{}}99\textsuperscript{th} Percentile \\ Execution \\Speedup\end{tabular}} \\ \hline
\multicolumn{8}{c}{\textbf{Rainbowcake Applications}} \\ \hline 
Dna-visualisation (R-DV) & NumPy & Scientific Computing & C3 & 2.30$\times$ & 2.26$\times$ & 2.03$\times$ & 1.99$\times$ \\  
Graph-bfs (R-GB) & igraph & Graph Processing & C1 & 1.71$\times$ & 1.66$\times$ & 1.55$\times$ & 1.54$\times$ \\  
Graph-mst (R-GM) & igraph & Graph Processing & C1 & 1.74$\times$ & 1.70$\times$ & 1.67$\times$ & 1.64$\times$ \\  
Graph-pagerank (R-GPR) & igraph & Graph Processing & C1 & 1.70$\times$ & 1.62$\times$ & 1.69$\times$ & 1.64$\times$ \\  
Sentiment-analysis (R-SA) & nltk, TextBlob & Natural Language Processing & C1 & 1.35$\times$ & 1.33$\times$ & 1.37$\times$ & 1.34$\times$ \\ \hline
\multicolumn{8}{c}{\textbf{FaaSLight Applications}} \\ \hline
Price-ml-predict (FL-PMP) & SciPy & Machine Learning & C1 & 1.31$\times$ & 1.30$\times$ & 1.37$\times$ & 1.36$\times$ \\  
Skimage-numpy (FL-SN) & SciPy & Image Processing & C1 & 1.41$\times$ & 1.36$\times$ & 1.41$\times$ & 1.37$\times$ \\  
Predict-wine-ml (FL-PWM) & pandas & Machine Learning & C1 & 1.76$\times$ & 1.68$\times$ & 1.59$\times$ & 1.52$\times$ \\  
Train-wine-ml (FL-TWM) & pandas & Machine Learning & C1 & 1.79$\times$ & 1.50$\times$ & 1.72$\times$ & 1.46$\times$ \\   
Sentiment-analysis (FL-SA) & pandas, SciPy & Natural Language Processing & C1 & 2.01$\times$ & 2.01$\times$ & 2.15$\times$ & 2.15$\times$ \\ \hline
\multicolumn{8}{c}{\textbf{FaaS Workbench Applications}} \\ \hline
Chameleon (FWB-CML) & pkg\_resources & Package Management & C3 & 1.17$\times$ & 1.05$\times$ & 1.24$\times$ & 1.07$\times$ \\  
Model-training (FWB-MT) & SciPy & Machine Learning & C1 & 1.21$\times$ & 1.09$\times$ & 1.20$\times$ & 1.09$\times$ \\  
Model-serving (FWB-MS) & SciPy & Machine Learning & C1 & 1.23$\times$ & 1.10$\times$ & 1.22$\times$ & 1.10$\times$ \\ \hline
\multicolumn{8}{c}{\textbf{Real-World Applications}} \\ \hline
OCRmyPDF & pdfminer & Document Processing & C4 & 1.42$\times$ & 1.19$\times$ & 1.63$\times$ & 1.00$\times$ \\ \hline
CVE-bin-tool & xmlschema & Security & C2 & 1.27$\times$ & 1.20$\times$ & 1.08$\times$ & 1.01$\times$ \\ \hline
\end{tabular}
\caption{Summary of performance improvement}
\label{table:summary}
\end{table*}

To report the results, we consider both the average and the 99\textsuperscript{th}-percentile latency as performance metrics. Reporting the 99\textsuperscript{th}-percentile latency is important because it captures the performance experienced by the slowest 1\% of requests, providing insights into potential outliers and bottlenecks. This metric ensures a comprehensive understanding of an application's performance,  helping to meet Service Level Agreements (SLAs) by ensuring that even under peak load conditions, the user experience remains consistent and reliable. Additionally, since serverless applications mostly use a pay-per-use billing model, it is crucial to use optimal memory resources. Therefore, we also report detailed memory optimization achieved through our evaluation.

\begin{figure}
    \setlength{\belowcaptionskip}{-13.9pt}
    \includegraphics[width=\linewidth]{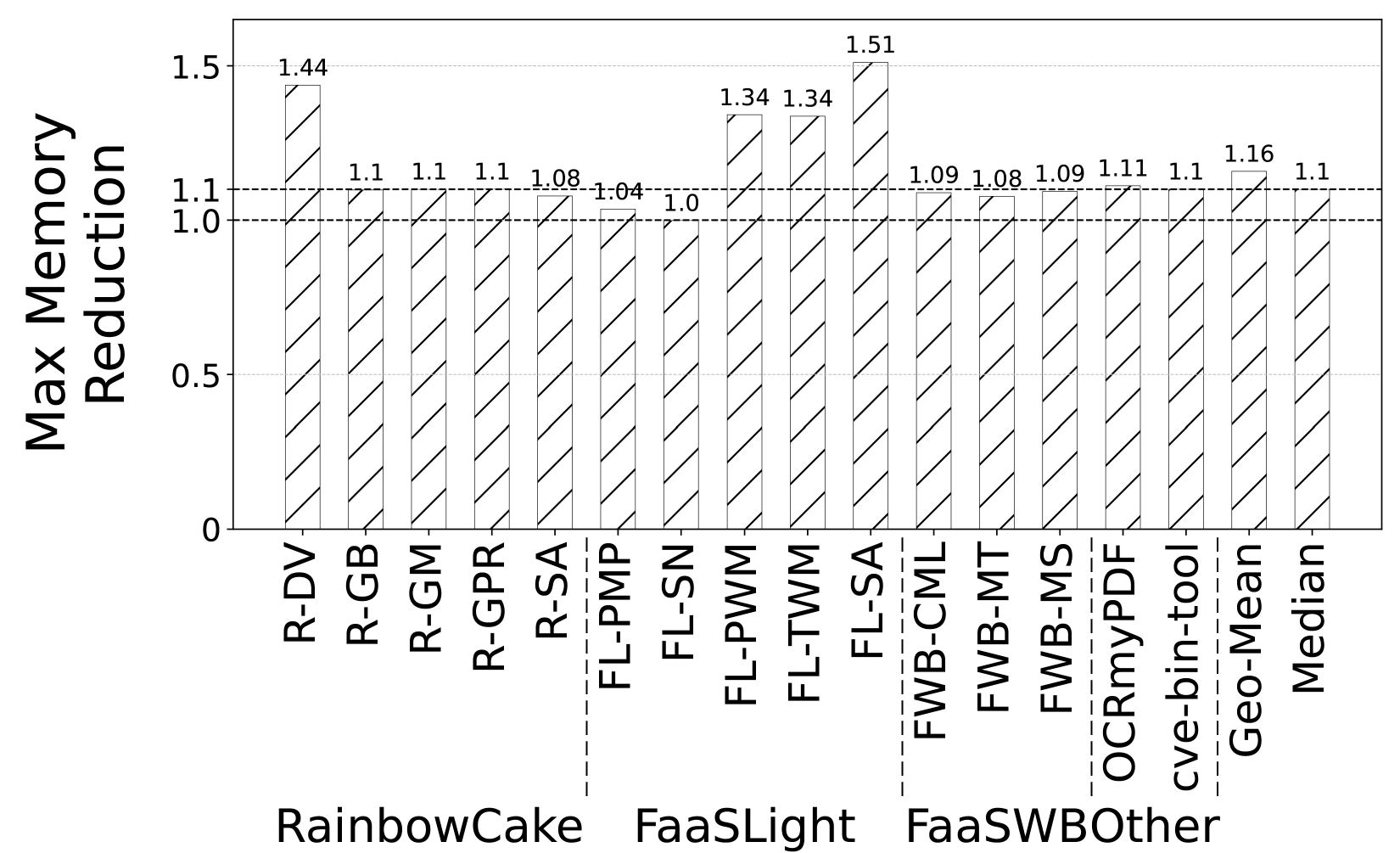}
    \caption{Memory reduction after \tool{} guided optimization.}
    \label{fig:mem_optimization}
\end{figure}

\paragraph{\textbf{Summary of evaluation}}
Table~\ref{table:summary} summarizes the inefficiencies detected by \tool{} along with the optimizations achieved after addressing them. Guided by \tool{}, our evaluation identifies inefficiencies in 15 out of 20 serverless applications evaluated. To assess the effectiveness of \tool{}'s inefficiency reports, we optimize these applications. The \tool{}-guided optimizations lead to improvements of up to 2.30$\times$ in initialization latency and 2.26$\times$ in execution latency. Figure~\ref{fig:mem_optimization} details the memory reductions achieved through these optimizations, showing up to a 1.51$\times$ reduction in memory usage. These results demonstrate the effectiveness of \tool{} and directly address evaluation question \textbf{Q2}.

\paragraph{\textbf{\tool{} vs FaaSLight}}
To better understand the effectiveness of \tool{}'s dynamic analysis capabilities in comparison to a static analysis-based approach, we examine the speedup achieved through \tool{}-guided optimization against the speedups reported by FaaSLight. Our comparison includes the FaaSLight applications optimized during our evaluation of \tool{}. Due to the inability to execute optimized FaaSLight applications directly, we rely on speedup data presented in the FaaSLight paper~\cite{liu2023faaslight}. Table~\ref{table:comparison} summarizes the results, demonstrating that \tool{}-guided optimization results in greater overall speedups in total response time and provides more significant reductions in memory usage.

\begin{table}[h!]
\centering
\fontsize{8pt}{9pt}\selectfont
\vspace{-0.2cm}
\begin{tabular}{@{}ccccc@{}}
\hline
\textbf{App ID} & \textbf{Tool} & \textbf{Version} & \begin{tabular}[c]{@{}c@{}}\textbf{Runtime} \\ \textbf{memory (MB)}\end{tabular} & \begin{tabular}[c]{@{}c@{}}\textbf{Total response} \\ \textbf{latency (ms)}\end{tabular} \\ \hline

\multirow{4}{*}{\begin{tabular}[c]{@{}c@{}}App4 \\scikit \\assign\end{tabular}} & \multirow{2}{*}{\begin{tabular}[c]{@{}c@{}}FaasLight \\ (Reported)\end{tabular}} & before & 142 & 4,534.38 \\ 
 &  & after & 140 ($1.01\times$) & 4,004.10 ($1.13\times$) \\ \cline{2-5}
 & \multirow{2}{*}{\begin{tabular}[c]{@{}c@{}}\tool{} \\ (Measured)\end{tabular}} & before & 123.64 & 3,184.67 \\ 
 &  & after & 119.38 ($1.04\times$) & 2,452.59 ($1.30\times$) \\ \hline

\multirow{4}{*}{\begin{tabular}[c]{@{}c@{}}App7 \\skimage \\lambda\end{tabular}} & \multirow{2}{*}{\begin{tabular}[c]{@{}c@{}}FaasLight \\ (Reported)\end{tabular}} & before & 228 & 7,165.54 \\ 
 &  & after & 130 ($1.75\times$) & 4,152.73 ($1.73\times$) \\ \cline{2-5}
 & \multirow{2}{*}{\begin{tabular}[c]{@{}c@{}}\tool{} \\ (Measured)\end{tabular}} & before & 112.09 & 1,821.73 \\ 
 &  & after & 112.21 ($1.00\times$) & 1,342.48 ($1.36\times$) \\ \hline

\multirow{4}{*}{\begin{tabular}[c]{@{}c@{}}App9 \\train wine \\ml-lambda\end{tabular}} & \multirow{2}{*}{\begin{tabular}[c]{@{}c@{}}FaasLight \\ (Reported)\end{tabular}} & before & 230 & 9,035.39 \\ 
 &  & after & 216 ($1.06\times$) & 7,470.49 ($1.21\times$) \\ \cline{2-5}
 & \multirow{2}{*}{\begin{tabular}[c]{@{}c@{}}\tool{} \\ (Measured)\end{tabular}} & before & 251.91 & 5,154.34 \\ 
 &  & after & 187.76 ($1.34\times$) & 3,059.18 ($1.68\times$) \\ \hline

\multirow{4}{*}{\begin{tabular}[c]{@{}c@{}}App9 \\predict wine \\ml-lambda\end{tabular}} & \multirow{2}{*}{\begin{tabular}[c]{@{}c@{}}FaasLight \\ (Reported)\end{tabular}} & before & 230 & 8,291.80 \\ 
 &  & after & 215 ($1.07\times$) & 7,071.03 ($1.17\times$) \\ \cline{2-5}
 & \multirow{2}{*}{\begin{tabular}[c]{@{}c@{}}\tool{} \\ (Measured)\end{tabular}} & before & 252.08 & 6,201.17 \\ 
 &  & after & 188.57 ($1.34\times$) & 4,147.68 ($1.50\times$) \\ \hline

\multirow{4}{*}{\begin{tabular}[c]{@{}c@{}}App11 \\sentiment \\analysis\end{tabular}} & \multirow{2}{*}{\begin{tabular}[c]{@{}c@{}}FaasLight \\ (Reported)\end{tabular}} & before & 182 & 5,551.03 \\ 
 &  & after & 141 ($1.29\times$) & 3,934.31 ($1.41\times$) \\ \cline{2-5}
 & \multirow{2}{*}{\begin{tabular}[c]{@{}c@{}}\tool{} \\ (Measured)\end{tabular}} & before & 203.54 & 4,331.43 \\ 
 &  & after & 134.72 ($1.51\times$) & 2,155.61 ($2.01\times$) \\ \hline
\end{tabular}
\caption{Comparison of \tool{} (Measured) vs FaasLight (Reported) metrics}
\label{table:comparison}
\end{table}
\vspace{-0.2cm}

\paragraph{\textbf{Overhead measurement}}
\label{overhead}
\begin{figure}
    \setlength{\belowcaptionskip}{-15pt}
    \includegraphics[width=0.9\linewidth]{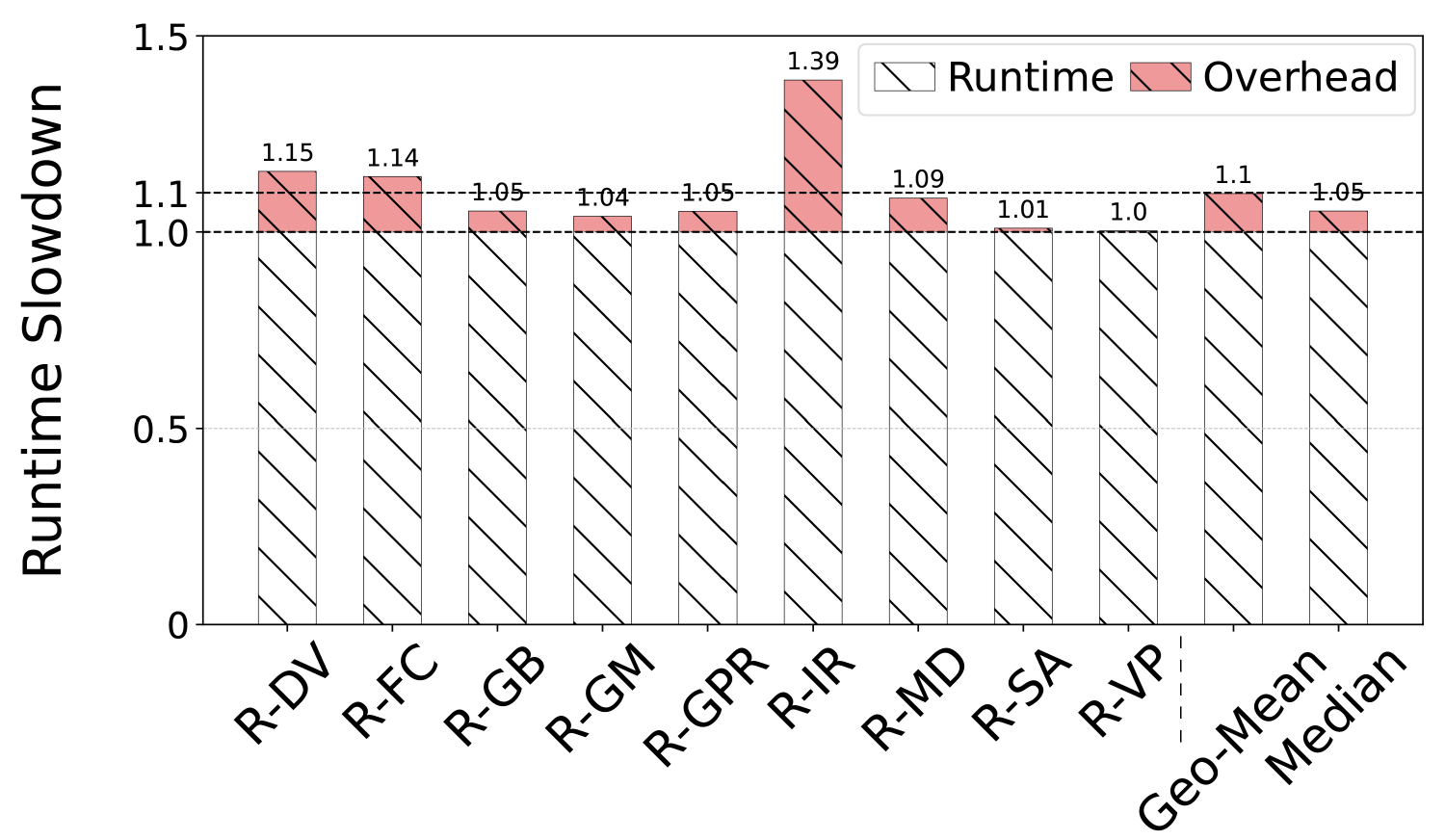}
    \caption{Runtime overhead of \tool{}.}
    \label{fig:overhead}
\end{figure}
To quantify the overhead incurred by \tool{} and answer \textbf{Q3}, we assess the runtime overhead of \tool{} on all nine Python serverless applications in the RainbowCake benchmark. During the assessment, each experiment is executed five times with 500 concurrent requests, comparing the runtime ratio with and without \tool{} monitoring enabled. The evaluation results are illustrated in Figure~\ref{fig:overhead}. The figure shows that most serverless applications experience a maximum overhead of 10\%. Additionally, \tool{} provides an API to configure the sampling rate.
\section{Case Studies}
This section discusses the detailed utility of \tool{} and attempts to answer evaluation question \textbf{Q1}. We present four case studies to demonstrate how \tool{} identifies inefficiencies and illustrate the accuracy of \tool{} in detecting these inefficiencies. Then, we discuss how we optimize each of the applications by utilizing the insights from the comprehensive inefficiency reports generated by \tool{}.

\subsubsection{\textbf{CVE Binary Analyzer}}

\texttt{CVE Binary Analyzer}, developed by Intel, detects system vulnerabilities using the NVD CVE list. Version 3.3 supports 360 vulnerabilities and 11 language-specific checkers, scanning binaries to report known vulnerabilities. It is often integrated with CI/CD systems, such as GitHub Actions, for automatic CVE scanning.


\paragraph{\textbf{The Scenario}} 
During the evaluation, we select 342 of the most popular GitHub repositories, each with a maximum archive size of 100 KB. We deploy the \texttt{CVE Binary Analyzer} as a Lambda function on AWS, integrate it with \tool{}, and use it to scan these repositories for vulnerabilities with the \texttt{OpenSSL} checker. \texttt{OpenSSL} is a popular library for secure communications, and it has had numerous vulnerabilities reported over time. By enabling the \texttt{OpenSSL} checker, the \texttt{CVE Binary Analyzer} scans the repositories to identify the version of \texttt{OpenSSL} in use and cross-references it against known CVEs associated with \texttt{OpenSSL}. We initiate 500 concurrent requests to the Lambda function, and \tool{} identifies inefficiencies in Python library usage within the \texttt{CVE Binary Analyzer}.

\paragraph{\textbf{\tool{} Insight}}
Table~\ref{table:cvebin_report} shows \tool{}'s evaluation report of the \texttt{CVE Binary Analyzer} application. \tool{} lists the Python libraries imported by \texttt{CVE Binary Analyzer} along with their overhead in initialization latency and utilization data. The \tool{} visualizer shows the call path to the library package, allowing users to view its corresponding usage.

Our evaluation finds that the \texttt{xmlschema} library accounts for more than 8.27\% of the total initialization latency, despite its utilization metrics showing it is sampled as little as 0.78\% of the total samples. A manual investigation reveals that the \texttt{xmlschema} library is only required when the \texttt{CVE Binary Analyzer} encounters a repository containing XML files with a Software Bill of Materials (SBOM) as the root.

\paragraph{\textbf{The Optimization}}
We optimize the \texttt{CVE Binary Analyzer} application by implementing lazy loading of the \texttt{xmlschema} library. This optimization results in a 1.27$\times$ improvement in initialization latency and a 1.20$\times$ improvement in execution latency. Additionally, lazy loading \texttt{xmlschema} reduces the maximum memory usage of \texttt{CVE Binary Analyzer} by 1.21$\times$.

\subsubsection{\textbf{DNA Visualization (R-DV)}}

\texttt{R-DV} is a DNA visualization application we evaluate as part of the RainbowCake benchmark. It is a scientific application designed to help researchers, scientists, and bioinformaticians to visualize, analyze, and interpret DNA sequences. \texttt{R-DV} reads from multiple DNA sequences, transforms the sequences, combines the transformed data, and generates visualization results. 

\paragraph{\textbf{The Scenario}} 
For evaluation, we deploy \texttt{R-DV} as a Lambda function on AWS. We use AWS Lambda ephemeral storage to store the DNA sequence data of the input genes and the visualization results from \texttt{R-DV}. Concurrent requests to the DNA visualization application are made using a command-line HTTP client. The handler function collects DNA sequences from the input gene files and uses version 0.3.1 of the \texttt{squiggle} library to transform these sequences into output visualizations.

\paragraph{\textbf{\tool{} Insight}}
Table~\ref{table:dnaviz_report} shows \tool{}'s evaluation report of the \texttt{R-DV} application. \tool{} lists the Python libraries imported by \texttt{R-DV}, along with the overhead they impose on initialization latency and their utilization metrics.

We observe that the \texttt{numpy} library accounts for over 63\% of the initialization time, despite its low utilization of 2.6\%. Our manual investigation reveals that \texttt{numpy} is used by the \texttt{squiggle} library but can be replaced by Python's standard runtime libraries. Further investigation confirms that the \texttt{numpy} dependency has also been removed in the latest version of \texttt{squiggle}.

\paragraph{\textbf{The Optimization}}
By replacing \texttt{numpy} with Python's standard runtime library, we optimize the \texttt{R-DV} application. This optimization leads to a 2.30$\times$ improvement in initialization latency and a 2.26$\times$ improvement in execution latency. Replacing \texttt{numpy} also reduces the maximum memory usage of \texttt{R-DV} by 1.44$\times$.

\begin{table}[h]
\fontsize{7.5pt}{8.5pt}\selectfont
\vspace{-0.2cm}
\begin{tabular}{cllrr}
\hline
\rowcolor{gray!30}
\multicolumn{5}{|c|}{\textbf{\tool{} Summary}} \\ \hline
\multicolumn{5}{|l|}{\textbf{Application:} cve\_binary\_analyzer} \\ \hline
\rowcolor{gray!30}
\multicolumn{1}{|l|}{} & \multicolumn{1}{l|}{\textbf{Package}} & \multicolumn{1}{l|}{\textbf{Util.}} & \multicolumn{1}{c|}{\textbf{\begin{tabular}[c]{@{}c@{}}Init.\\ Overhead\end{tabular}}} & \multicolumn{1}{r|}{\textbf{File}} \\ \hline
\rowcolor{red!30}
\multicolumn{1}{|c|}{+} & \multicolumn{1}{l|}{xmlschema} & \multicolumn{1}{r|}{0.78} & \multicolumn{1}{r|}{8.27} & \multicolumn{1}{r|}{../xmlschema/\_\_init\_\_.py} \\ \hline
\multicolumn{1}{|c|}{+} & \multicolumn{1}{l|}{elementpath} & \multicolumn{1}{r|}{1.48} & \multicolumn{1}{r|}{8.17} & \multicolumn{1}{r|}{../elementpath/\_\_init\_\_.py} \\ \hline
\multicolumn{1}{|c|}{} & \multicolumn{1}{l|}{...} & \multicolumn{1}{r|}{...} & \multicolumn{1}{r|}{...} & \multicolumn{1}{r|}{...} \\ \hline
\rowcolor{gray!30}
\multicolumn{5}{|c|}{\textbf{Call Path}} \\ \hline
\rowcolor{gray!30}
\multicolumn{1}{|l|}{\textbf{}} & \multicolumn{1}{l|}{\textbf{Package}} & \multicolumn{3}{l|}{\textbf{Path}} \\ \hline
\multicolumn{1}{|l|}{-} & \multicolumn{1}{l|}{xmlschema} & \multicolumn{3}{l|}{\begin{tabular}[c]{@{}l@{}}handler.py:11\\ \quad-> cve\_bin\_tool/cli.py:71\\ \quad\quad->cve\_bin\_tool/sbom\_detection.py:8\\ 
\quad\quad\quad-> cve\_bin\_tool/validator.py:11\end{tabular}} \\ \hline
\multicolumn{1}{l}{} &  &  & \multicolumn{1}{l}{} & \multicolumn{1}{l}{}
\end{tabular}
\caption{\tool{} report for CVE binary analyzer}
\label{table:cvebin_report}
\end{table}

\begin{table}[h]
\fontsize{7.5pt}{8.5pt}\selectfont
\vspace{-0.2cm}
\begin{tabular}{cllrr}
\hline
\rowcolor{gray!30}
\multicolumn{5}{|c|}{\textbf{\tool{} Summary}} \\ \hline
\multicolumn{5}{|l|}{\textbf{Application:} rainbowcake\_dna\_visualization} \\ \hline
\rowcolor{gray!30}
\multicolumn{1}{|l|}{} & \multicolumn{1}{l|}{\textbf{Package}} & \multicolumn{1}{l|}{\textbf{Util.}} & \multicolumn{1}{c|}{\textbf{\begin{tabular}[c]{@{}c@{}}Init.\\ Overhead\end{tabular}}} & \multicolumn{1}{r|}{\textbf{File}} \\ \hline
\rowcolor{red!30}
\multicolumn{1}{|c|}{+} & \multicolumn{1}{l|}{numpy} & \multicolumn{1}{r|}{2.60} & \multicolumn{1}{r|}{63.27} & \multicolumn{1}{r|}{../numpy/\_\_init\_\_.py} \\ \hline
\multicolumn{1}{|c|}{+} & \multicolumn{1}{l|}{urllib3} & \multicolumn{1}{r|}{0.29} & \multicolumn{1}{r|}{10.07} & \multicolumn{1}{r|}{../urllib3/\_\_init\_\_.py} \\ \hline
\multicolumn{1}{|c|}{} & \multicolumn{1}{l|}{...} & \multicolumn{1}{r|}{...} & \multicolumn{1}{r|}{...} & \multicolumn{1}{r|}{...} \\ \hline
\rowcolor{gray!30}
\multicolumn{5}{|c|}{\textbf{Call Path}} \\ \hline
\rowcolor{gray!30}
\multicolumn{1}{|l|}{\textbf{}} & \multicolumn{1}{l|}{\textbf{Package}} & \multicolumn{3}{l|}{\textbf{Path}} \\ \hline
\multicolumn{1}{|l|}{-} & \multicolumn{1}{l|}{numpy} & \multicolumn{3}{l|}{\begin{tabular}[c]{@{}l@{}}handler.py:8\\ \quad-> squiggle/\_\_init\_\_.py:1\\
\quad\quad-> squiggle/squiggle.py:1\end{tabular}} \\ \hline
\multicolumn{1}{l}{} &  &  & \multicolumn{1}{l}{} & \multicolumn{1}{l}{}
\end{tabular}
\caption{\tool{} report for DNA Visualization (R-DV)}
\label{table:dnaviz_report}
\end{table}
\vspace{-0.2cm}

\subsubsection{\textbf{Sentiment Analysis (R-SA)}}

\texttt{R-SA} is a sentiment analysis application featured in the RainbowCake benchmark. This scientific tool processes human-readable text, analyzes its sentiment, and classifies the text as positive, negative, or neutral. We assess \texttt{R-SA} as part of our evaluation.

\paragraph{\textbf{The Scenario}}
After integrating \tool{}, we upload the \texttt{R-SA} application package and its dependencies to an AWS S3 bucket. Then, we deploy \texttt{R-SA} as a Lambda function by transmitting the source code from S3. The Lambda handler reads the text input and utilizes version 3.8.1 of the natural language toolkit \texttt{nltk} library for tokenization and the \texttt{TextBlob} library for sentiment analysis. \texttt{R-SA} reports the sentiment analysis results in terms of subjectivity and polarity scores. To generate the inefficiency report, we execute 500 concurrent requests to the corresponding \texttt{R-SA} Lambda function.

\paragraph{\textbf{\tool{} Insight}}
Table~\ref{table:rsa_report} presents the inefficiency report for the \texttt{R-SA} application generated by \tool{}. We discovered that although the \texttt{nltk} library has a low utilization of 5.33\%, it contributes to around 70\% of the initialization latency. Further analysis revealed that the sub-modules \texttt{sem, stem, parse, tag} of the \texttt{nltk} library add 25\% overhead to the initialization time, yet they are never used by \texttt{R-SA}.

\paragraph{\textbf{The Optimization}}
Based on the guidance from \tool{}'s report, we optimize the \texttt{R-SA} application by lazy loading the sub-modules \texttt{nltk.sem, nltk.stem, nltk.parse, nltk.tag}. This optimization results in a 1.35$\times$ improvement in initialization latency and a 1.33$\times$ improvement in execution latency. Additionally, it reduces the maximum memory usage of \texttt{R-SA} by 1.07$\times$.

\break

\begin{table}[h]
\fontsize{7.5pt}{8.5pt}\selectfont
\begin{tabular}{cllrr}
\hline
\rowcolor{gray!30}
\multicolumn{5}{|c|}{\textbf{\tool{} Summary}} \\ \hline
\multicolumn{5}{|l|}{\textbf{Application:} rainbowcake\_sentiment\_analysis.json} \\ \hline
\rowcolor{gray!30}
\multicolumn{1}{|l|}{} & \multicolumn{1}{l|}{\textbf{Package}} & \multicolumn{1}{l|}{\textbf{Util.}} & \multicolumn{1}{c|}{\textbf{\begin{tabular}[c]{@{}c@{}}Init.\\ Overhead\end{tabular}}} & \multicolumn{1}{r|}{\textbf{File}} \\ \hline
\rowcolor{red!10}
\multicolumn{1}{|c|}{-} & \multicolumn{1}{l|}{nltk} & \multicolumn{1}{r|}{5.33} & \multicolumn{1}{r|}{69.93} & \multicolumn{1}{r|}{../nltk/\_\_init\_\_.py} \\ \hline
\rowcolor{red!30}
\multicolumn{1}{|c|}{+} & \multicolumn{1}{l|}{nltk.sem} & \multicolumn{1}{r|}{0} & \multicolumn{1}{r|}{8.25} & \multicolumn{1}{r|}{nltk/sem/\_\_init\_\_.py} \\ \hline
\multicolumn{1}{|c|}{} & \multicolumn{1}{l|}{...} & \multicolumn{1}{r|}{...} & \multicolumn{1}{r|}{...} & \multicolumn{1}{r|}{...} \\ \hline
\rowcolor{gray!30}
\multicolumn{5}{|c|}{\textbf{Call Path}} \\ \hline
\rowcolor{gray!30}
\multicolumn{1}{|l|}{\textbf{}} & \multicolumn{1}{l|}{\textbf{Package}} & \multicolumn{3}{l|}{\textbf{Path}} \\ \hline
\multicolumn{1}{|l|}{-} & \multicolumn{1}{l|}{nltk.sem} & \multicolumn{3}{l|}{\begin{tabular}[c]{@{}l@{}}handler.py:2\\ \quad-> nltk/\_\_init\_\_.py:147\\
\quad\quad-> <... parent path ...>\\
\quad\quad\quad-> nltk/sem/\_\_init\_\_.py:44\end{tabular}} \\ \hline
\multicolumn{1}{l}{} &  &  & \multicolumn{1}{l}{} & \multicolumn{1}{l}{}
\end{tabular}
\caption{\tool{} report for Sentiment Analysis (R-SA)}
\label{table:rsa_report}
\end{table}
\vspace{-0.2 cm}

\begin{table}[h]
\fontsize{7.5pt}{8.5pt}\selectfont
\begin{tabular}{cllrr}
\hline
\rowcolor{gray!30}
\multicolumn{5}{|c|}{\textbf{\tool{} Summary}} \\ \hline
\multicolumn{5}{|l|}{\textbf{Application:} model\_training.json} \\ \hline
\rowcolor{gray!30}
\multicolumn{1}{|l|}{} & \multicolumn{1}{l|}{\textbf{Package}} & \multicolumn{1}{l|}{\textbf{Util.}} & \multicolumn{1}{c|}{\textbf{\begin{tabular}[c]{@{}c@{}}Init.\\ Overhead\end{tabular}}} & \multicolumn{1}{r|}{\textbf{File}} \\ \hline
\rowcolor{red!10}
\multicolumn{1}{|c|}{-} & \multicolumn{1}{l|}{scipy} & \multicolumn{1}{r|}{5.94} & \multicolumn{1}{r|}{37.92} & \multicolumn{1}{r|}{../scipy/\_\_init\_\_.py} \\ \hline
\rowcolor{red!30}
\multicolumn{1}{|c|}{+} & \multicolumn{1}{l|}{scipy.stats} & \multicolumn{1}{r|}{0} & \multicolumn{1}{r|}{13.25} & \multicolumn{1}{r|}{../scipy/\_\_init\_\_.py} \\ \hline
\multicolumn{1}{|c|}{} & \multicolumn{1}{l|}{...} & \multicolumn{1}{r|}{...} & \multicolumn{1}{r|}{...} & \multicolumn{1}{r|}{...} \\ \hline
\rowcolor{gray!30}
\multicolumn{5}{|c|}{\textbf{Call Path}} \\ \hline
\rowcolor{gray!30}
\multicolumn{1}{|l|}{\textbf{}} & \multicolumn{1}{l|}{\textbf{Package}} & \multicolumn{3}{l|}{\textbf{Path}} \\ \hline
\multicolumn{1}{|l|}{-} & \multicolumn{1}{l|}{scipy.stats} & \multicolumn{3}{l|}{\begin{tabular}[c]{@{}l@{}}lambda\_function.py:5\\ 
\quad-> sklearn/\_\_init\_\_.py:87\\
\quad\quad-> <... parent path ...>\\ 
\quad\quad\quad-> scipy/status/\_\_init\_\_.py:605\end{tabular}} \\ \hline
\multicolumn{1}{l}{} &  &  & \multicolumn{1}{l}{} & \multicolumn{1}{l}{}
\end{tabular}
\caption{\tool{} report for Model Training (FWB-MT)}
\label{table:mt_report}
\end{table}
\vspace{-0.6 cm}

\subsubsection{\textbf{Machine Learning Model Training and Model Serving}}

We evaluate the \texttt{FWB-MT} and \texttt{FWB-MS} applications from the FaaSWorkbench benchmark. The \texttt{FWB-MT} application trains a machine learning model, while \texttt{FWB-MS} uses the pre-trained model for prediction.

\paragraph{\textbf{The Scenario}} 
The \texttt{FWB-MT} Lambda reads 548K Amazon food reviews, trains a logistic regression model and stores it in S3. The \texttt{FWB-MS} retrieves the model, predicts review sentiment, and returns the prediction and latency as a JSON response. We invoke each lambda function 500 times to generate a profiling report.

\paragraph{\textbf{\tool{} Insight}}
Table~\ref{table:mt_report} presents the inefficiency report for the \texttt{FWB-MT} application. The \texttt{FWB-MS} application uses the same dependencies, and \tool{} generated a similar report. We discovered that although the \texttt{scipy.stats} library has no utilization, it contributes to over 13.25\% of the initialization latency in the \texttt{FWB-MT} application and 12.22\% in the \texttt{FWB-MS} application.

\paragraph{\textbf{The Optimization}}
Following \tool{}'s recommendations, we optimize \texttt{FWB-MT} and \texttt{FWB-MS} by lazy loading \texttt{scipy.stats}. This reduces \texttt{FWB-MT}'s initialization latency by 1.21$\times$, execution latency by 1.09$\times$, and memory usage by 1.08$\times$. \texttt{FWB-MS} sees a 1.23$\times$ improvement in initialization latency, 1.10$\times$ in execution latency, and 1.09$\times$ in memory usage.
\section{Discussion}
\paragraph{\textbf{Implications}}
During our study, we found that newer versions of libraries often incur higher initialization overheads than their predecessors due to escalating complexity over years of development. Tools like \tool{} can help developers pinpoint areas for improvement. Despite efforts by the system community to devise strategies to reduce cold-start time, inefficient code will continue to impact latency-sensitive applications with suboptimal resource usage. Adopting efficient coding practices with tool support, such as that offered by \tool{}, can help mitigate these issues and optimize resource utilization.
\paragraph{\textbf{Limitations}}
\tool{} has two limitations: First, the current implementation of \tool{} is written in Python and incurs higher overhead at high-frequency sampling. To reduce the overhead, we will implement \tool{} in the native language, similar to other state-of-the-art Python profilers. Second, \tool{} cannot monitor native code execution. As a glue language, Python often relies on libraries written in native code. However, current \tool{} relies on the Python library to unwind the call stack and reconstruct the call paths. As a result, \tool{} does not collect the call path of native code execution. We will address the issue by collecting native call paths and reconstructing joint native and Python call paths implemented in other Python profilers.

\section{Threats to validity}
The effectiveness of \tool{}'s optimization guidance depends on the inputs provided to serverless functions during profiling. Different inputs can lead to varying execution paths and library usage. Furthermore, as a statistical profiler, \tool{} relies on the Law of Large Numbers, requiring many samples over time for accuracy. To address this, we tested \tool{} with standard inputs and collected numerous samples by monitoring many serverless function invocations.
\section{Conclusions}

This paper proposed \tool{}, a profiler designed to detect inefficient library usage in serverless Python applications. \tool{} requires no modification to the runtime, or privileged access, and it can identify inefficiencies with minimal overhead of 10\%. During the evaluation, \tool{} effectively identified inefficiencies in 15 serverless applications. \tool{} profile guided optimization, results up to 2.30$\times$ speedup in initialization latency, 2.26$\times$ speedup in the cold-start execution time as well as a 1.51$\times$ reduction in memory usage. \tool{} provides serverless application developers with valuable insights, ensuring efficient utilization of Python libraries.
\bibliographystyle{ACM-Reference-Format}
\bibliography{BibFile}
\end{document}